\documentclass[aps,showpacs,preprint]{revtex4}

\usepackage{graphicx,dcolumn,array,bm,amsmath,appendix,lineno}
\usepackage{epstopdf}
\usepackage{subfigure}
\usepackage{CJK}
\usepackage[normalem]{ulem}
\usepackage{amsmath}
\usepackage{mathrsfs}
\usepackage{supertabular}
\usepackage{textcomp}
\usepackage{cancel}
\usepackage{xcolor}

\def\bk{{\mbox{\boldmath $k$}}}
\def\nab{{\mbox{\boldmath $\nabla$}}}
\newcommand{\bnll}[1]{\begin{subequations}\label{#1}\begin{eqnarray}}
\newcommand{\enll}{\end{eqnarray}\end{subequations}}

\def\bk{{\mbox{\boldmath $k$}}}

\def\br{{\mbox{\boldmath $r$}}}

\def\bbk{{\mbox{\boldmath $\scriptstyle k$}}}

\def\nab{{\mbox{\boldmath $\nabla$}}}
\makeatletter

\newcommand{\Rmnum}[1]{\expandafter@slowromancap\romannumeral #1@}
\makeatother

\begin{document}

\title{Three-Dimensional Approach Applied to Quasi-stationary States of Deformed $\alpha$-Emitters}
\author{Ruijia Li }
\affiliation{School of Physics, Nanjing University, Nanjing 210093, China}

\author{Chang Xu }
\email{cxu@nju.edu.cn}
\affiliation{School of Physics, Nanjing University, Nanjing 210093, China}
\affiliation{Institute for Nonperturbative Physics, Nanjing University, Nanjing 210093, China}

\begin{abstract}
We apply a three-dimensional (3D) approach to investigate the quasi-stationary states of well-deformed $\alpha$-emitters. With a splitting of the anisotropic 3D potential into internal and external parts at a separation surface, the 3D $\alpha$-cluster decay width is determined by the initial wave function of a true bound state of an anisotropic harmonic oscillator potential and a non-resonance scattering wave function of Coulomb potential. Substantial difference between the 1D and 3D decay width is found for typical $\alpha$-emitters with large quadrupole and hexadecapole deformations.
\end{abstract}

\pacs{23.60. + e, 21.60.Gx}

\maketitle

\section{Introduction}
\label{s1}

The quantum tunneling problem of $\alpha$-cluster decay provides insight into not only the quasi-stationary states of unstable nuclei but also the nature of $\alpha$-clustering in surrounding nuclear medium \cite{DLW,Delion09,Buck,cxu06,Mohr06,review1,review2,Ismail16,Perez19,Denisov09,Royer,Poenaru}. Much of our understanding of $\alpha$-cluster decay comes from the well-established quantum picture where an $\alpha$-cluster tunnels through the Coulomb barrier after its formation on the surface of nuclei. The $\alpha$-clustering process is rather complex to handle, which in principle involves a quantum four-body problem embedded in nuclear medium \cite{THSR,quartet2017,Po,Yasu21,Eiji20,SY20,SY21,Xu2017,Xu2016}. In contrast, the tunneling probability can be quantitatively estimated from the WKB approximation by assuming an $\alpha$-cluster interacting with a spherical daughter nucleus. This assumption is appropriate for $\alpha$-emitters with small deformations \cite{Buck,cxu06,Mohr06}. However, large deformations could be involved in the $\alpha$-cluster decay of heavy nuclei and exotic nuclei far away from the $\beta$-stability line \cite{Oganessian,Batchelder,Te2019,Zhang}. The exact 3D decay width of well-deformed $\alpha$-emitters is difficult to obtain by treating self-consistently both deformation and mixing of angular momenta. An empirical way to include effect of large deformations is to average the tunneling probabilities of all angles, which yields an enhanced $\alpha$-cluster tunneling probability. For $\alpha$-transitions with measured core excitation energies, the total wave function of the system can be expanded in terms of the ground and excited states channels and the coupled Schrödinger equations can be solved \cite{Barmore,channel1,Delion06}.

The exact solution of quasi-stationary state with an anisotropic potential is quite interesting in many quantum systems \cite{QT,Spins,Electrons,QD,Superconductor,TPA,MTPA}. In this work, we apply a 3D approach to investigate the quasi-stationary state of $\alpha$-cluster decay problem, namely the 3D two-potential approach (3D-TPA) \cite{MDTPA}. We emphasize that the quasi-stationary state of $\alpha$-cluster decay highly resembles a bound state more than a scattering state. The anisotropic $\alpha$-core potential is properly divided into inner and outer potentials on a separation surface inside the Coulomb barrier. The formed $\alpha$-cluster is considered to stay initially in the bound state generated by the inner potential, and then transforms to a quasi-stationary state by switching on the outer potential. Using the time-dependent perturbation theory, both the decay width and the energy shift of $\alpha$-cluster decay can be well defined by inner 3D bound state wave function and outer scattering state wave function on the separation surface. The multi-dimensional quantum tunneling problem for well-deformed $\alpha$-emitters is reduced to a problem of true bound state plus non-resonance scattering state \cite{MDTPA}. It is worth noting that the choice of separation surface does not affect the final results as long as it is inside the classical forbidden region. The challenge here is the numerical solution of the inner 3D bound state wave function on the separation surface, whose value is approximately of the order of 10$^{-15}$. Although several numerical methods can be applied to solve the 3D Schrödinger equation to obtain the bound state wave function such as the grid-based approach, the imaginary time propagation method, and the basis expansion method, however, it is still an open question to obtain accurately the inner 3D bound state wave function at large distances due to the limitation of matrix size or the number of bases \cite{Imaginary,Gaussian}. In this sense, the anisotropic harmonic oscillator potential with exact solutions is a good choice to simulate the internal $\alpha$-core 3D potential. The 1D and 3D decay widths are compared for several typical non-spherical $\alpha$-emitters with large quadrupole and hexadecapole deformations. Note that the formation process of the $\alpha$-cluster on nuclear surface is not touched here, which in principle does not affect the comparison between 1D and 3D decay widths.

The paper is organized as follows: In Sec.\ref{s2}, we give the formalism of 3D two-potential approach for the multi-dimensional quantum tunneling problem. The anisotropic $\alpha$-core effective potential and the choice of separation surface are introduced in Sec.\ref{s3}. In Sec.\ref{s4}, both the inner 3D bound state wave function and outer scattering state wave function are explicitly given. In Sec.\ref{s5}, the 3D approach is applied to several deformed $\alpha$-emitters $^{236}\textrm{U}$, $^{244}\textrm{Pu}$, $^{246}\textrm{Cm}$, $^{250}\textrm{Cf}$, and $^{254}\textrm{Fm}$. The possible theoretical uncertainties are also discussed. Conclusions are presented in Sec.\ref{s6}.

\section{Formalism of 3D two-potential approach for multi-dimensional quantum tunneling problem}
\label{s2}
The time-dependent perturbation theory is applied to solve the multi-dimensional tunneling problems in 3D-TPA \cite{MDTPA}. The main integrant is the separation of the 3D $\alpha$-core potential $V(\br)$ into the inner $U(\br)$ and outer $W(\br)$ potentials on the separation surface $\mathcal{S}$
\begin{equation}
    V(\br )=U(\br )+W(\br ).
\end{equation}
The inner potential is
\begin{equation}
 U(\br )=\left\{
    \begin{aligned}
      &V(\br )& \text{inner region}\\
       &U_0& \text{outer region},
    \end{aligned}
   \right.
   \label{cut1}
\end{equation}
where $U_0$ is the minimal value of $V(\br )$ on the separation surface $\mathcal{S}$. The outer potential is
\begin{equation}
 W(\br )=\left\{
    \begin{aligned}
      &0&\text{inner region}\\
       &V(\br )-U_0&\text{outer region}.
    \end{aligned}
   \right.
     \label{cut2}
\end{equation}
The quasi-stationary state of $\alpha$-cluster decay can be described by the time-dependent Schr\"odinger equation,
\begin{equation}
\begin{aligned}
  i\hbar{\partial\over\partial t}|\Psi(t)\rangle &=\bigg[-\frac{\hbar^2}{2\mu}\nab^2+V(\br )\bigg]|\Psi(t)\rangle\\
  &=\bigg[H_0+W(\br)\bigg]|\Psi(t)\rangle,
\label{nn1}
\end{aligned}
\end{equation}
where $H_0=-\frac{\hbar^2}{2\mu}\nab^2+U(\br )$ is the Hamiltonian of bound state $|\Phi _i\rangle$ confined in the inner potential $U(\br)$, and the corresponding time-independent Schr\"odinger equation is $[-\frac{\hbar^2}{2\mu}\nab^2+U(\br )]|\Phi_i\rangle=E_0|\Phi_i\rangle$. The separation surface $\mathcal{S}$ is not necessarily spherical, as long as it is taken between the equipotential surfaces $S_1$ and $S_2$. At $t>0$, the ``unperturbed'' bound state $|\Phi _i\rangle$ is no longer an eigenstate of the total Hamiltonian $H=H_0+W(\br)$, but a wave packet spreading in time due to the perturbation $W(\br)$
\begin{equation}
   |\Psi(t)\rangle=b_0(t)|\Phi_i\rangle e^{-iE_0t/\hbar}
   +\int
b_{\textbf{k}}(t)|\varphi_{\textbf{k}}\rangle e^{-iE_{\textbf{k}}t/\hbar}\frac{d\textbf{k}}{(2\pi)^3},
\label{b2}
\end{equation}
where $b_0(t)$ and $b_{\textbf{k}}(t)$ are the probability amplitudes of finding the system in the eigenstates $|\Phi_i\rangle$ and $|\varphi_{\textbf{k}}\rangle$, respectively. The amplitudes $b_0(t)$ and $b_{\textbf{k}}(t)$ can be found from Eq.(\ref{nn1}) with the initial condition: $b_0(t)=1, b_{\textbf{k}}(t)=0$. The energy shift $Re(\epsilon_0)$ and the width $\Gamma=-2Im(\epsilon_0)$ of the quasi-stationary state are directly related to the pole in the complex E-plane using the Green’s function technique \cite{TPA}
\begin{equation}
\epsilon_0=E-E_0=\langle\Phi_i|W|\Phi_i\rangle+\langle\Phi_i|W\tilde G(E)W|\Phi_i\rangle.
\label{bb7}
\end{equation}
The Green's function $\tilde G$ is given by
\begin{equation}
\tilde G(E)=G_0(E)\Big[1+\tilde W\tilde G(E)\Big],
\label{bb5}
\end{equation}
where $\tilde W=W+U_0$ is used instead of $W$, to ensure the potential vanishes for $r\rightarrow \infty$. $G_0(E)$ is given by
\begin{equation}
G_0(E)=\frac{1-\Lambda}{E+U_0-K-U(\br)},\qquad
\Lambda =|\Phi_i\rangle \langle\Phi_i|.
\label{b6}
\end{equation}

Note that the above derivations are general, but the numerical solution of $\tilde G$ is rather difficult and converges very slowly. To make it feasible within the capacity of computer calculation, $\tilde G$ is expanded in powers of $G_{\tilde W}$, namely the Green function corresponding to $\tilde W(\br)$
\begin{equation}
   \tilde G(E)=G_{\tilde W}(E)+G_{\tilde W}(E)(U-U_0)\tilde G(E)
-\tilde G_{\tilde W}(E)\Lambda
\Big[1+\tilde W\tilde G(E)\Big].
\label{bb9}
\end{equation}
Substitute Eq.(\ref{bb9}) into Eq.(\ref{bb7}), then retain the first order and assume that the energy shift is small compared to $E_0$, one can replace $G_{\tilde W}(E)$ by $G_{\tilde W}(E_0)$
\begin{equation}
E=E_0+\langle\Phi_i|W|\Phi_i\rangle
+\langle\Phi_i|W G_{\tilde W}(E_0)W|\Phi_i\rangle\,.
\label{bb11}
\end{equation}
The Schr\"odinger equation of the Green's function is
\begin{equation}
\left [E_0-K-\tilde W(\br )\right ]
G_{\tilde W}(E_0;\br ,\br' )=\delta (\br -\br' ),
\label{b12}
\end{equation}
and the spectral representation for the Green's function is
\begin{equation}
G_{\tilde W}=\int {|\varphi_{\textbf{k}}\rangle\langle \varphi_{\textbf{k}}|\over E_0-E_{\textbf{k}}-i\eta}
{d\textbf{k}\over (2\pi)^3},
\label{b17}
\end{equation}
where $E_{\textbf{k}}=\hbar^2\textbf{k}^2/2\mu$. The $|\varphi_{\textbf{k}}\rangle$ is the eigenstate corresponding to non-resonance scattering states, which satisfies the following Schr\"odinger equation
\begin{equation}
\Big[K+\tilde W(\br )\Big]|\varphi_{\textbf{k}}\rangle=E_{\textbf{k}}|\varphi_{\textbf{k}}\rangle.
\label{scatterings}
\end{equation}

One can obtain the total width $\Gamma$ as an integral over the partial width $\Gamma_{\textbf{k}}$ \cite{MDTPA}
\begin{equation}
\Gamma =\left. \int\Gamma_{\textbf{k} }{dk_1 dk_2\over (2\pi)^2}
\right |_{|\textbf{k} |=k_0},
\label{bb19}
\end{equation}
where $k_0=\sqrt{2mE_0/\hbar^2}$. The partial width $\Gamma_{\textbf{k}}$ is
\begin{equation}
\Gamma_{\bbk}=
{\hbar^2\over 4\mu k_3}\left |\int_{\br\in\mathcal{B}}\Phi_i(\br)
\stackrel{\leftrightarrow}\nab_{\textbf{r}}\varphi_{\textbf{k}}(\br)d\sigma
\right |^2_{|\textbf{k}|=k_0},
\label{bb20}
\end{equation}
where the symbol $\stackrel{\leftrightarrow}\nab_{\textbf{r}}$ means the gradient on the
right minus the gradient on the left. $\bk=\{k_1,k_2,k_3\}$ is the momentum vector. {We note that Eq.(\ref{bb20}) is similar to the Bardeen formula for the tunneling coupling between adjoining wells from a many-particle point of view, which is widely used in the solid state and atomic physics \cite{Bardeen,Gurvitz}.} Finally, the decay half-life is given by
\begin{equation}
    T_{1/2}=\frac{\hbar\ln{2}}{P_{\alpha}\Gamma},
\end{equation}
where P$_{\alpha}$ is the formation probability of $\alpha$-cluster on nuclear surface. Recent microscopic calculation of the $\alpha$-cluster formation probability for the ideal Po isotopes has been performed by the quartetting wave function approach (QWFA), in which the intrinsic motion of the four nucleons forming the $\alpha$-cluster and the center of mass motion between the $\alpha$-cluster and the core are correctly treated \cite{SY20,SY21,Xu2016,Xu2017}. Empirically, the $\alpha$-cluster formation probability $P_{\alpha}$ is known to change abruptly across the major shell closures and differs for even-even, odd-A and odd-odd nuclei, as indicated by the experimental systematics. The decay width $\Gamma$ can be directly obtained by using Eq.(\ref{bb19}) and Eq.(\ref{bb20}), however, it is more convenient to re-formulate $\Gamma$ in the following way
\begin{equation}
\begin{aligned}
    \Gamma=\frac{\hbar^2}{4\mu} \int_0^{2\pi}\int_0^\pi\frac{k_0 \sin{\theta_k} d\theta_k d\phi_k}{(2\pi)^2}\times\Bigg|\int_0^{2\pi}\int_0^\pi \Phi_i(r,\theta,\phi)
\stackrel{\leftrightarrow}\nab_{r}\varphi_{k}(r,\theta,\phi)
R_s(\theta,\phi)& \\
   \sqrt{[\frac{d R_s(\theta,\phi)}{d\phi}]^2+\sin{\theta}^2\bigg[\big[\frac{d R_s(\theta,\phi)}{d\theta}\big]^2+R_s^2(\theta,\phi)\bigg]}d\theta d\phi\Bigg|_{r=R_s(\theta,\phi)}^2,&
    \label{gam3}
\end{aligned}
\end{equation}
where $R_s(\theta,\phi)$ is the radius of the separation surface $\mathcal{S}$. For spherical emitters, Eq.(\ref{gam3}) reduces exactly to the 1D decay width $\Gamma={\hbar^2\over \mu k_0}|\phi_i(R)\chi_k'(R)-\chi_k(R)\phi_i'(R)|^2$ \cite{TPA}.

\section{$\alpha$-core 3D effective potential and separation surface}
\label{s3}
\begin{figure}[htb]
\begin{center}
\leavevmode
\includegraphics[width=0.7\textwidth]{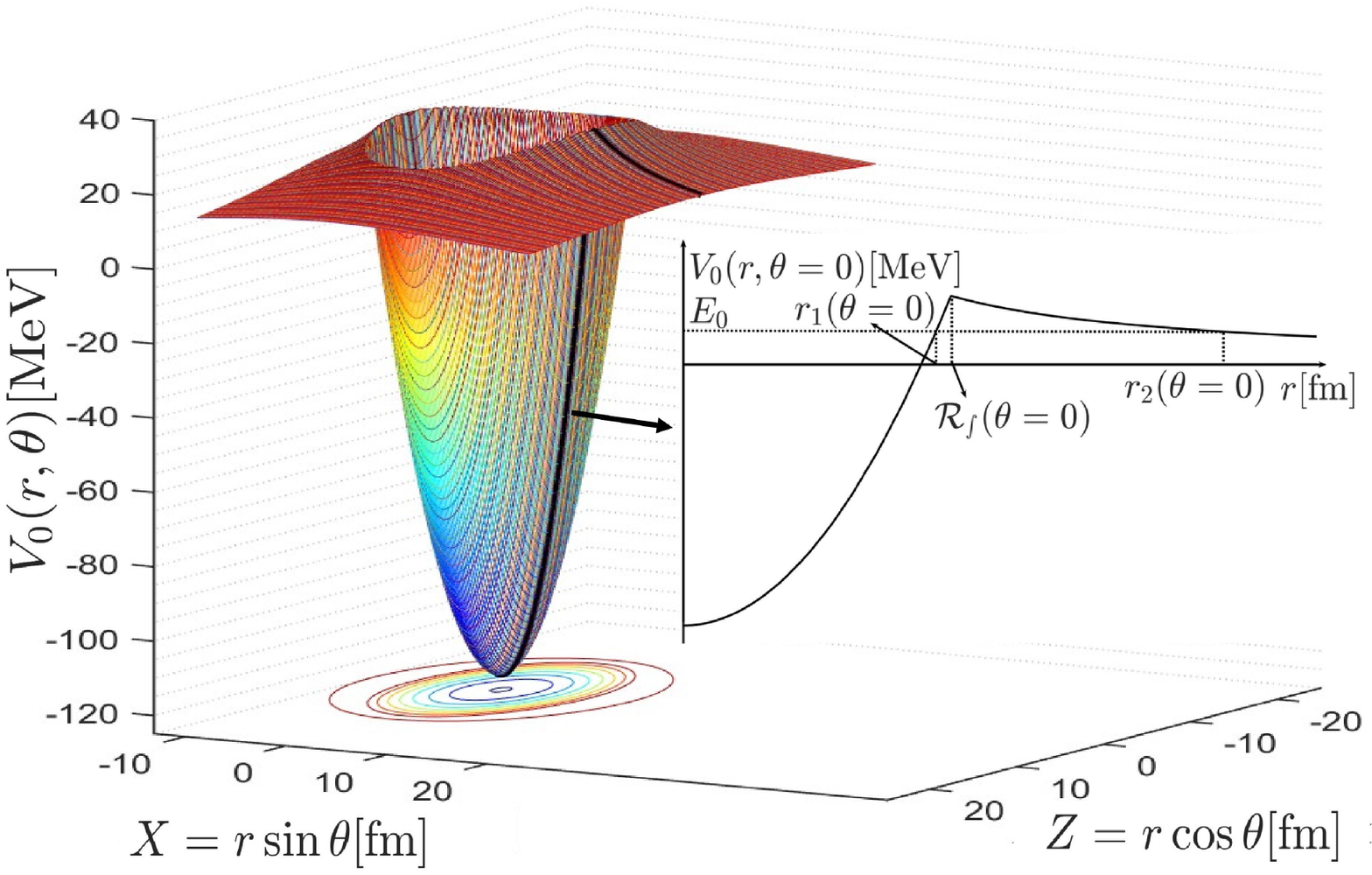}
\caption{The $\alpha$-core 3D potential $V_0(r,\theta)$ in Cartesian coordinate system. The inner part of $V_0(r,\theta)$ is an anisotropic harmonic oscillator potential, and the outer part of $V_0(r,\theta)$ a deformed Coulomb potential. {For demonstration, the $\alpha$-core potential $V_0(r,\theta)$ with the angle $\theta=0$ is shown in the small panel. $r_1(\theta=0)$ and $r_2(\theta=0)$ are the first and second turning points, respectively. $\mathcal{R_s}(\theta=0)$ is the matching point where $V_I(r,\theta=0)-\kappa\omega_0\hbar\big[L^2-\langle L^2\rangle_N\big]=V_C(r,\theta=0)$.}
\label{figrr2}}
\end{center}
\end{figure}
The 3D-TPA requires the information on 3D bound-state wave function deep inside the classical forbidden region where the wave function decreases exponentially and is extremely small (of the order of ~10$^{-15}$). This poses a big challenge for numerical calculations, even with high-performance parallel computing. Here we simulate the 3D $\alpha$-core potential with an anisotropic harmonic oscillator potential that is of rotational symmetry. The relevant Hamiltonian is of Nilsson-form, which has a $L^2$ correction term \cite{Nilsson,SPM}
\begin{equation}
    H=-\frac{\hbar^2}{2\mu}\nab^2+V_0(r,\theta),\
\end{equation}
with
 \begin{equation}
  V_0(r,\theta)=\left\{
    \begin{aligned}
      &V_I(r,\theta)-\kappa\omega_0\hbar\bigg[L^2-\langle L^2\rangle_N\bigg]&&r\leq \mathcal{R_s}(\theta)\\
       &V_C(r,\theta)&&r > \mathcal{R_s}(\theta),
    \end{aligned}
   \right.
\end{equation}
where {$\mathcal{R_s}(\theta)$ describes the boundary condition $V_I(r,\theta)-\kappa\omega_0\hbar\big[L^2-\langle L^2\rangle_N\big]=V_C(r,\theta)$}. The anisotropic harmonic oscillator potential with depth parameter $D$ is
\begin{equation}
    V_I(r,\theta)= \frac{1}{2}m \omega^2\bigg[(r\cos\theta)^2+\gamma^2(r\sin\theta)^2\bigg]-D,
\end{equation}
where $\omega$, $\omega_0$ and $\gamma$ read
\begin{equation}
\omega=\sqrt{\frac{2D}{\mu\Big[R(0)/a_r\Big]^2}},~~~~\omega_0=\sqrt{\frac{2D}{\mu\Big[R_0/a_r\Big]^2}},~~~~ \gamma=\frac{R(0)}{R(\frac{\pi}{2})}.
\end{equation}
The deformed Coulomb potential is
\begin{equation}
\begin{aligned}
      V_C(r,\theta)&=\frac{1.44 Z_c Z_d }{r}\bigg[1+\frac{3}{5}(\frac{R_0}{r})^2\beta_2\sum_m Y_{2m}(\theta)\bigg]\\
    &+\sqrt{\frac{1 8}{7\pi}}(\frac{R_0}{r})^2\bigg[-\beta_2^2 Y_{20}(\theta)+\frac{3}{4}\sqrt{5}(\frac{R_0}{r})^2\beta_2^2 Y_{40}(\theta)\bigg],
\end{aligned}
\end{equation}
The half-density radius $R(\theta)$ is given by
\begin{equation}
  R(\theta)=R_0\bigg[1+\beta_2 Y_{20}(\theta)+\beta_4 Y_{40}(\theta)\bigg],
\end{equation}
where the parameter $R_0=1.07A_d^{1/3} $\cite{rrr}, and $\beta_2$ and $\beta_4$ are quadrupole and hexadecapole deformations, respectively. The deformed Coulomb potential $V_C$ can be regarded as an isotropic potential at very large distances. The $L^2$ correction term is introduced to eliminate the angular momentum degeneracy, resulting in different depths of potential for different angular momenta.

\begin{figure}[htb]
\begin{center}
\leavevmode
\includegraphics[width=0.7\textwidth]{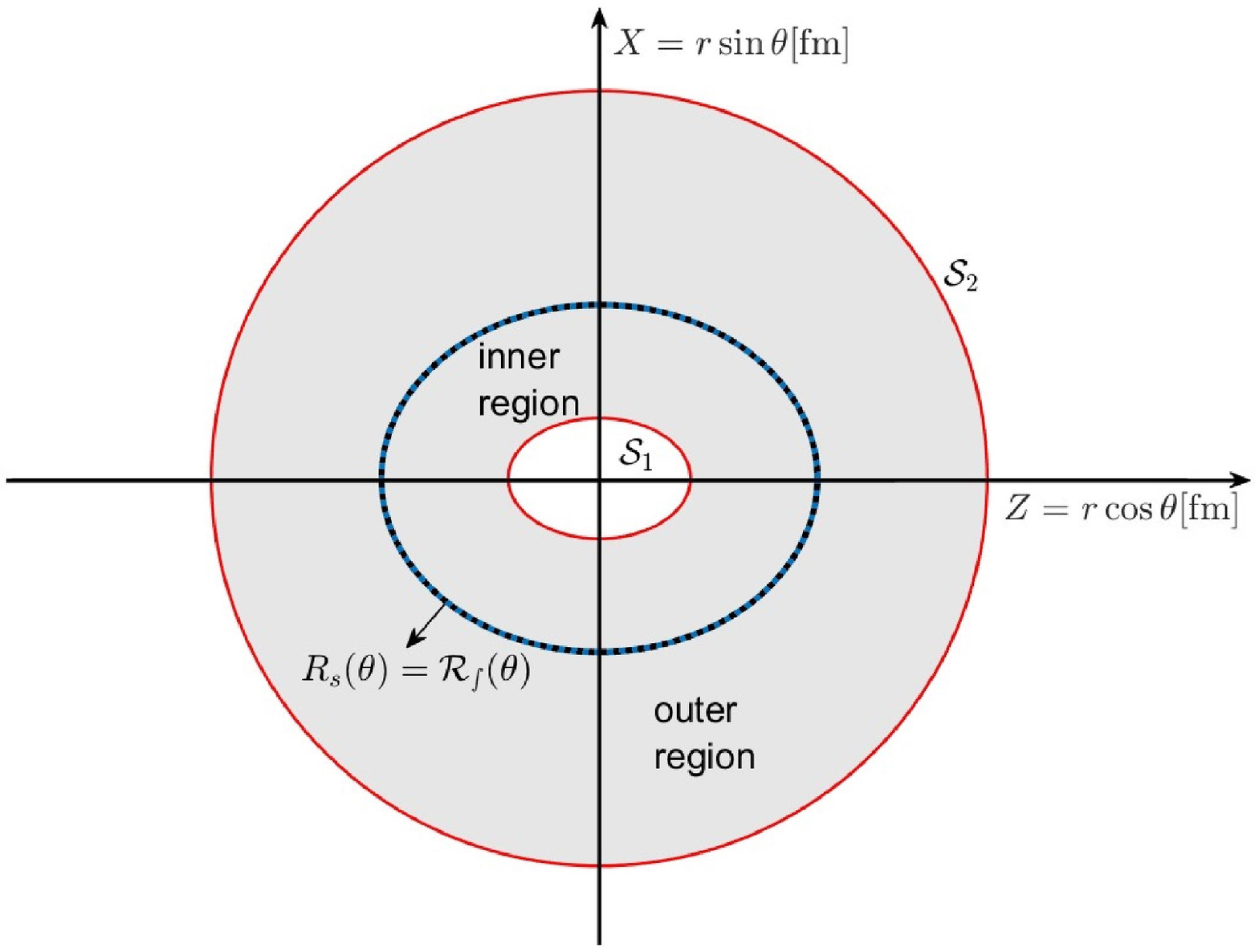}
\caption{The separation of the $\alpha$-core potential $V_0(r,\theta)$ into inner and outer regions. $S_1$ and $S_2$ denote the  equipotential surfaces with the same energy $E_0$. {The separation surface $\mathcal{S}$, denoted by the radius $R_s(\theta)$, is chosen as coinciding with $\mathcal{R_s}(\theta)$}.
\label{figr3}}
\end{center}
\end{figure}

{The details of the $\alpha$-core 3D potential $V_0(r,\theta)$ are shown in Fig.\ref{figrr2}, in which the inner potential joins with the outer Coulomb potential at $\mathcal{R_s}(\theta)$. The total $\alpha$-core potential is divided into two parts by the separation surface $\mathcal{S}$, which can be chosen between the equipotential surfaces $S_1$ and $S_2$ with the same energy $E_0$. Here, we take the separation surface $\mathcal{S}$ as coinciding with $\mathcal{R_s}(\theta)$ (see Fig.\ref{figr3}). In this way, one can minimize the numerical errors associated with $\Phi_i(r,\theta,\phi)$ and $\varphi_{\textbf{k}}(r,\theta,\phi)$. Note that the boundary $\mathcal{R_s}(\theta)$ is azimuth-independent ($\phi$-independent). Thus, the separation surface $\mathcal{S}$ is also azimuth-independent in our calculations.}

\section{Inner 3D bound state wave function and outer scattering state wave function}
\label{s4}
We use the analytical 3D wave function of anisotropic harmonic oscillator potential as an approximation of $\Phi_i(r,\theta,\phi)$, which is the eigenstate of $E_0$ corresponding to the inner potential. As a matter of fact, we only need the information of $\Phi_i(r,\theta,\phi)$ on the surface $\mathcal{S}$ where the 3D wave function of anisotropic harmonic oscillator is considered to be a good approximation for well-deformed $\alpha$-emitters
\begin{equation}
\Phi_i(r,\theta,\phi)=\Psi_{n_{\rho}}^m(r,\theta)\Psi_{n_z}(r,\theta)\frac{e^{i m\phi}}{\sqrt{2\pi}},
 \label{phi1}
 \end{equation}
 where
 \begin{equation}
 \begin{aligned}   \Psi_{n_{\rho}}^m(r,\theta)&=N_{n_{\rho}}^m\beta_{\perp}^{|m|+1}\sqrt{2} (r\sin{\theta})^{|m|} e^{-(r^2\sin{\theta}^2\beta_{\perp}^2)/2}
    L_{n_{\rho}}^{|m|}(r\sin{\theta}),
 \end{aligned}
 \end{equation}
 and
 \begin{equation}
 \Psi_{n_z}(r,\theta)=N_{n_z}\beta_z^{1/2}e^{-r^2\cos{\theta}^2/2}H_{n_z}(r\cos\theta),
  \end{equation}
with parameters $\beta_{\perp}$, $\beta_z$, $N_{n_{\rho}}^m$ and $N_{n_z}$
 \begin{subequations}
 \begin{align}
       \beta_{\perp}=(\frac{m\omega\gamma}{\hbar})^{1/2}&,\quad \beta_z=(\frac{m\omega}{\hbar})^{1/2},\\
        N_{n_{\rho}}^m=(\frac{n_{\rho}!}{(n_{\rho}+|m|)!})&,\quad
    N_{n_z}=(\frac{1}{\sqrt{\pi}2^{n_z}n_z!})^{1/2}.\quad
 \end{align}
 \end{subequations}

\begin{figure}[htb]
\begin{center}
\leavevmode
\includegraphics[width=0.7\textwidth]{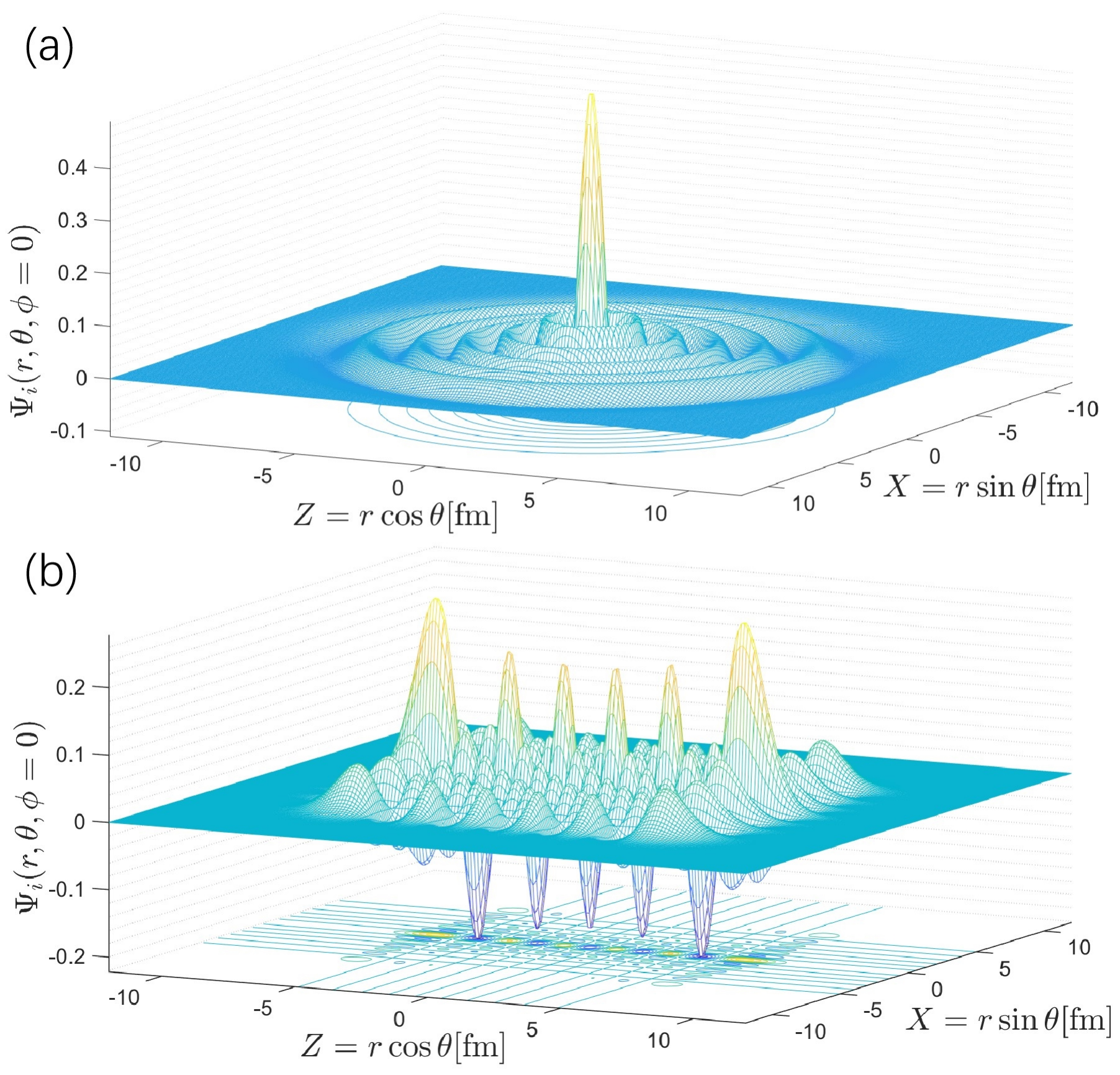}
\caption{(a) The 3D inner wave function $\Phi_i(r,\theta,\phi=0)$ of the spherical case with $\beta_2=0$ and $\beta_4=0$. (b)  $\Phi_i(r,\theta,\phi=0)$ of the well-deformed case with $\beta_2=0.226$, $\beta_4=0.108$. For any $\phi$, the $\Phi_i(r,\theta,\phi)$ has the same shape.
\label{fig8r}}
\end{center}
\end{figure}

\begin{figure}[htb]
\begin{center}
\leavevmode
\includegraphics[width=0.7\textwidth]{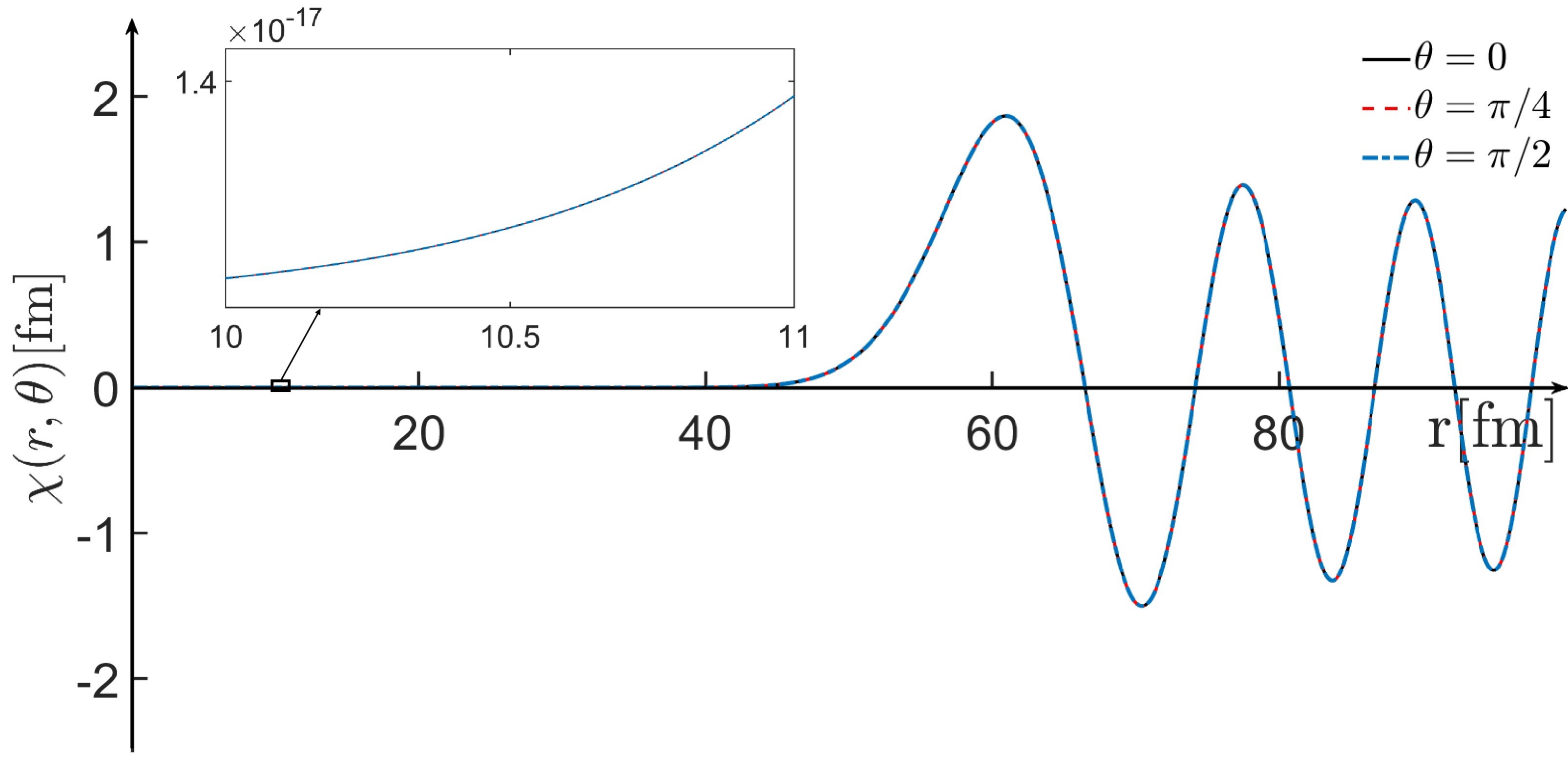}
\caption{The wave function $\chi(r,\theta)$ in the outer potential $\tilde W (r,\theta)$. The black, blue, and red curves refer to $\chi(r,\theta)$ with $\theta=0$, $\theta=\pi/4$ and $\theta=\pi/2$, respectively, which almost coincide with each other.
\label{fig10}}
\end{center}
\end{figure}

The expectation value $E_0$ is given by \cite{SPM}
 \begin{equation}
 \begin{aligned}
    E_0= \langle n_z n_{\rho}& m|H|n_z n_{\rho} m\rangle=(2n_{\rho}+m+1)\hbar\omega\gamma+(n_z+\frac{1}{2})\hbar\omega\\
     &+\hbar\omega_0\kappa\big[\frac{1}{2}(2n_z-G-\frac{1}{2})^2-m^2-\frac{1}{8}\big]-D,
     \label{expectation}
 \end{aligned}
 \end{equation}
where the depth $D$ can be obtained by matching the value of experimental decay energy $Q_{\alpha}$ with $E_0$. The choice of $n_{\rho}$ and $n_z$ should fulfill the so-called Wildermuth-Tang rule \cite{WT}
\begin{equation}
    G=2n_{\rho}+n_z=\sum_{i=1}^4 g_i,
\end{equation}
where $G$ is the global quantum number and $g_i$ are the oscillator quantum numbers of the nucleons forming the $\alpha$-cluster. The $G$ value is restricted by the Pauli principle. For instance, $G$ is usually taken as $G=22$ for heavy nuclei with neutron numbers $N>126$. In principle, $n_z$ can be taken as an even number from 0 to $G$. Here, $n_z$ is taken as $n_z=10$ as an example.

The scattering wave function $\varphi_k(r,\theta,\phi)$ is solved from the scattering Schr\"odinger equation Eq.(\ref{scatterings}) with the outer potential (see Fig.3(b)), which has the form \cite{scattering,QM}
\begin{equation}
\varphi_k(r,\theta,\phi)=2\pi\sum_{l,m} \sqrt{\frac{2}{\pi}} i^l e^{\pm i\sigma_l} Y_{lm}(\theta,\phi) Y_{lm}^*(\theta_k,\phi_k)\frac{\chi(r,\theta)}{kr},
    \label{comps1}
\end{equation}
where $\chi(r,\theta)$ can be represented by the linear combination of the regular $F_l(r)$ and irregular $G_l(r)$ solutions of Coulomb potential. The corresponding coefficients in the linear combination of $F_l(r)$ and $G_l(r)$ are directly related to the non-resonant scattering phase shift for the outer potential $\tilde W(r)$ \cite{MTPA}
\begin{equation}
\chi(r,\theta)=\cos{\delta_l(\theta)}F_l(r)+\sin{\delta_l(\theta)}G_l(r),
\end{equation}
where the relative phase $\delta_l$ obeys
\begin{equation}
   \tan{\delta_l}(\theta)=\frac{-F_lR_s'(\theta)+\alpha F_lR_s(\theta)}{G_lR_s'(\theta)-\alpha G_lR_s(\theta)},
\end{equation}
and
\begin{equation}
     \alpha=\frac{\sqrt{2\mu(U_0-Q_{\alpha})}}{\hbar}.
\end{equation}
The inner wave function $\Phi_i(r,\theta,\phi)$ and the outer Coulomb function $\chi(r,\theta)$ are shown in Fig.\ref{fig8r} and Fig.\ref{fig10}, respectively. For spherical emitters, the inner wave function shown in Fig.\ref{fig8r}(a) is the spherical harmonic oscillator wave function. For well-deformed emitters, the inner wave function is approximated by the anisotropic harmonic oscillator wave function as shown in Fig.\ref{fig8r}(b). As for the scattering state, the wave function $\chi(r,\theta)$ becomes almost isotropic at large distances (see Fig.\ref{fig10}).

\section{Comparison between 1D and 3D cases and uncertainty analysis}
\label{s5}
As shown in Table \ref{tab},  all the $\alpha$-cluster emitters selected in calculations are well-deformed with $\beta_2>0.22$ and far away from the major shell closures $Z=82$ and $N=126$. Moreover, only the  $\alpha$-transitions between ground-states ($0^+ \longrightarrow 0^+$) are considered in order to check the validity of 3D-TPA. There are three adjustable parameters in 3D-TPA that should be determined, namely $a_r$, $\kappa$, and $D$. $a_r$ is associated with the angular frequency $\omega$ of harmonic oscillator. $\kappa$ is the strength coefficient of orbit correction term. $D$ is the depth of harmonic oscillator potential, which can be obtained by matching the experimental decay energy $Q_{\alpha}$ with the energy $E_0$ once the quantum numbers $G,n_z,m$ are determined. It is noted that different quantum numbers $G,n_z,m$ can be chosen. Here the same parameters and quantum numbers are used for all emitters, namely, $a_r=0.7883, \kappa=0.0054, G=22, n_z=10$, and $m=0$.
\\
\begin{table}[htb]
    \centering
     \caption{Comparison between 1D and 3D $\alpha$-decay half-lives (in log base 10 and in seconds) for $^{254}$Fm, $^{250}$Cf, $^{246}$Cm, $^{244}$Pu, and $^{236}$U. The information on both parent and daughter nuclei are listed in columns 1–4. Column 5 gives the experimental $\alpha$-decay energy $Q_{\alpha}$. The theoretical quadrupole and the hexadecapole deformations are listed in columns 6 and 7, respectively. The experimental $\alpha$-decay half-lives are given in column 8. In the last two columns, the 1D and 3D $\alpha$-cluster decay half-lives from TPA are given.}
\setlength{\tabcolsep}{3.5mm}
    \begin{tabular}{cccccccccc}
    \hline
    $A_p$ & $Z_p$  & $A_d$ & $ Z_d $ & $Q_{\alpha}(MeV)$& $\beta_2$ & $\beta_4$&$T_{Exp.}$&$T_{Cal.}^{1D}$&$T_{Cal.}^{3D}$\\
       \hline\hline
236 & U & 232 &Th&4.572&0.226&0.108&15.003&16.2416 &14.9246\\
244 & Pu & 240 &U&4.6655&0.237&0.061&15.5015&16.5278 &15.5721\\
246 & Cm & 242 &Pu&5.4748&0.249&0.051&11.2615&12.3441&11.2006\\
250 & Cf & 246 &Cm&6.1284&0.250&0.027&8.6984&9.32804&8.63641\\
254 & Fm & 250 &Cf&7.307&0.251&0.015&4.1415&4.73273&4.14148\\
\hline
    \end{tabular}
    \label{tab}
\end{table}
\begin{figure}[htb]
\begin{center}
\leavevmode
\includegraphics[width=0.7\textwidth]{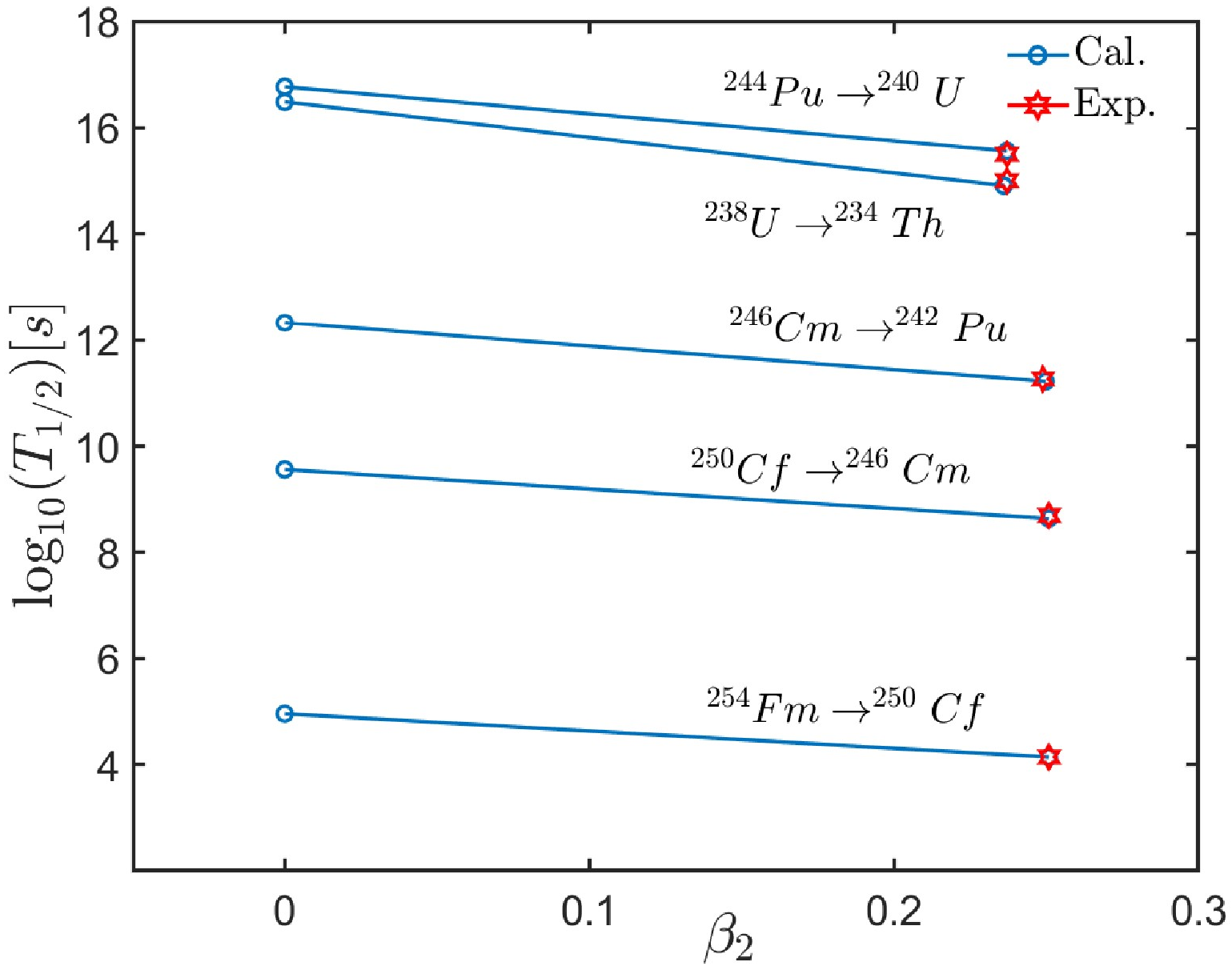}
\caption{The comparison of $\alpha$-decay half-lives (in logarithm with a base 10) in spherical and deformed cases. The red hexagrams refer to the experimental half-lives and the blue circles the calculated results from 3D-TPA.
\label{fig5r}}
\end{center}
\end{figure}

We show in Table \ref{tab} the comparison between 1D and 3D $\alpha$-decay half-lives for typical $\alpha$-emitter $^{236}\textrm{U}$, $^{244}\textrm{Pu}$, $^{246}\textrm{Cm}$, $^{250}\textrm{Cf}$, and $^{254}\textrm{Fm}$. Note that the quadrupole and hexadecapole deformations are taken from Ref.\cite{bb24}. $Q_{\alpha}$ and experimental results of $\alpha$-decay half-lives are taken from Refs.\cite{AME1,AME2}. One can see from Table \ref{tab} that the experimental $\alpha$-decay half-lives vary in a quite large range from $10^{4}$ to $10^{15}$ seconds. This is also helpful for testing the validity of present 3D model for not only short-lived but also long-lived $\alpha$-emitters. The formation problem of $\alpha$-cluster on the surface of parent nucleus is not considered here and its value is assumed to be unity (P$_{\alpha}$=1). This assumption is reasonable because the $\alpha$-cluster formation probability is known to change dramatically in the vicinity of shell closures but smoothly in the open-shell region here. One can see from Table \ref{tab} that deviations of the calculated $\alpha$-decay half-lives exist between the 1D case and 3D case. For all the $\alpha$-emitters considered in this work, the largest deviation occurs for the decay of $^{236}$U, which is possibly due to its small decay energy. In general, the 1D results are all reduced by taking nuclear deformation into account (shown also in Fig.\ref{fig5r}). Similar conclusions can also be found in the empirical approaches \cite{cxu06}. More importantly, the 3D-TPA reproduces nicely the systematics of experimental data with only one set of parameters. This is quite different from the 1D-TPA without the consideration of deformation, in which one may need to adjust the parameters such as the depth of potential for each emitter.

Finally, we discuss the possible theoretical uncertainties of present 3D-TPA calculations, which mainly come from the following aspects: a) the error associated with the approximations such as the replacement of $G_{\tilde W}(E)$ by $G_{\tilde W}(E_0)$ in 3D-TPA. The correction due to this approximation is expected to be negligible because of very small energy shift from $E_0$ to $E$; b) the inner 3D wave function is approximated by the exact solutions of an anisotropic harmonic oscillator potential. We solved numerically the bound-state wave function corresponding to the inner potential in the spherical case, and found its value only deviates from the exact result by several percentages on the separation surface; c) the scattering wave function $\chi(r,\theta)$ is considered to be isotropic on the separation surface. We have checked this approximation by taking the decay of $^{236}U$ as an example, and found that the final results are almost not affected.

\section{Summary}
\label{s6}
Large deformation is relevant to the $\alpha$-cluster decay of heavy nuclei and nuclides far away from the $\beta$-stability line. In this work, we apply a three-dimensional approach with the Nilsson-form Hamiltonian to calculate the decay widths of typical $\alpha$-emitters $^{236}\textrm{U}$, $^{244}\textrm{Pu}$, $^{246}\textrm{Cm}$, $^{250}\textrm{Cf}$, and $^{254}\textrm{Fm}$ by dividing the 3D effective potential into a bound-state inner region and a scattering outer region.  The inner wave function can be well approximated by the exact solution of the anisotropic harmonic oscillator potential, and the scattering wave function can be safely considered as isotropic on the separation surface. Substantial difference is found between 1D and 3D decay width for favored transitions of these $\alpha$-emitters. The systematics of experimental $\alpha$-cluster decay half-lives is nicely reproduced. In the future, state-of-art numerical approaches can be applied to evaluate accurately the inner 3D wave function at large distances for arbitrary 3D potentials. Moreover, the combination of the present 3D-TPA and approaches of clustering such as quartetting wave function approach should be performed in order to predict reliably the $\alpha$-decay half-lives of unknown nuclei far away from the major shell closures.
\\
\\
$\bf{Acknowledgments}$

The work is supported by the National Natural Science Foundation of China (Grant No. 12275129) and the Fundamental Research Funds for the Central Universities (Grant No. 020414380209).

\end{document}